\documentclass[fleqn,10pt]{wlscirep}
\usepackage[utf8]{inputenc}
\usepackage[T1]{fontenc}
\usepackage{amsmath}

\usepackage{braket}

\def \pathpics {pictures}
\title{Entanglement of annihilation photons}

\author[1,*]{Alexander Ivashkin}
\author[1]{Dzhonrid Abdurashitov}
\author[1,2]{Alexander Baranov}
\author[1]{Fedor Guber}
\author[1]{Sergey Morozov}
\author[1,3,]{Sultan Musin}
\author[1,3,]{Alexander Strizhak}
\author[1]{Igor Tkachev}

\affil[1]{Institute for Nuclear Research RAS, Moscow, 117312, Russia}
\affil[2]{National Research Nuclear University MEPhI, Moscow, 115409, Russia}
\affil[3]{Moscow Institute of Physics and Technology, Moscow, 141701, Russia}

\affil[*]{ivashkin@inr.ru}

\begin{abstract}
We present  a new experimental study of the quantum entanglement of photon pairs produced in positron-electron annihilation at rest. Each annihilation photon has an energy that is five orders of magnitude higher than the energy of photons in optical experiments, which provides a unique opportunity for controlled formation of decoherent states and successive measurements of the same photon.The experimental setup includes a system of Compton polarimeters to measure the Compton scattering of annihilation photons in entangled and decoherent states. Decoherent states are prepared by pre-scattering of one of the initial photons prior to measurements in polarimeters. For the first time, a direct comparison of the polarization correlations of annihilation photons in the entangled and thus prepared separable states has been carried out.  The angular distributions of scattered photons turned out to be the same in both quantum states, which is an unexpected discovery for the quantum-entangled positron emission tomography.  Moreover, the correlation function in the Bell's inequality is also the same for entangled and separable states.  It follows that, despite numerous measurements in a series of experiments, there is still no experimental proof of the entanglement of annihilation photons.  These results are in line with recent theoretical predictions of an identical Compton scattering cross-section for entangled and specific mixed separable quantum states and cast doubt on the universality of the Bell's theorem for testing the entanglement and nonlocality in quantum theory.
\end{abstract}
\begin{document}

\flushbottom
\maketitle
%
%
\thispagestyle{empty}
\section*{Introduction}

The entanglement of a quantum system is a simple consequence of the superposition principle and means that the state of the system cannot be represented as a product of the states of individual subsystems. Historically, however, this term was introduced by Schrödinger, referring to the entanglement of our knowledge of quantum systems \cite{Schrodinger1, Schrodinger2}.
Such an interpretation of entanglement is quite relevant for the current situation with a system of two photons formed by positron-electron annihilation at rest.  The description of this system has been the subject of pioneering work on quantum entanglement in the past century and, as we will see below, is still unclear in some respects. 

The study of this two-photon system has a long and remarkable history, which can be divided into several stages. Initially, the idea of measuring a pair of annihilation photons was proposed in 1946 by Wheeler \cite{Wheeler}. He considered at that time a hypothetical bound system of an electron and a positron with an orbital moment equal to one or zero. In the latter case, due to the conservation of angular momentum and parity, the photons of the pair formed during positron-electron annihilation have mutually perpendicular polarization. In the same article, in order to test the predicted correlation of photon polarization, J. Wheeler proposed an experimental scheme with two Compton scatterers and detectors of scattered photons, which has already become a classic. Since photons are scattered predominantly perpendicular to the polarization plane, the dependence of the number of registered photons on the angle between the scattering planes should have a maximum or minimum at $90^\circ$ or $0^\circ$, respectively.

Almost simultaneously, two theoretical papers by Pryce and Ward \cite{Pryce} and by Snyder \textit{et al.} \cite{Snyder} predicted the behavior of the angular distribution of scattered annihilation gamma rays. These distributions were obtained on the basis of the Klein-Nishina  formula \cite{Nishina} with the following   initial state
\begin{equation}
 {\Psi} = \frac{1}{\sqrt2}( \ket{{H_1}{V_2}} + \ket{{V_1}{H_2}}),
 \label{eq:wavefunc}
\end{equation}
where H(V) represent the horizontal (vertical) linear polarization of the first or second photon.  In this state, photons do not have a definite polarization, albeit their polarizations are mutually orthogonal. It was shown that the ratio $R = {N_{\perp}}/{N_{\parallel}}$ of the number of counts for perpendicular and parallel orientations of scattering planes reaches a maximum of $R=2.85$ in the case when photons are scattered at an angle of $82^0$ to their initial momenta. These predictions were brilliantly confirmed in a pioneering experiment by Wu and Shaknov \cite{Wu}. Their experimental ratio $R=2.04\pm0.08$ was consistent with the theory, taking into account the finite solid angles of the detectors.  

 Seven years later, the results obtained were seriously rethought in the paper by Bohm and Aharonov \cite{Bohm}, devoted to the experimental verification of the famous EPR paradox by Einstein, Podolsky and Rosen \cite{EPR-paradox}.   The authors emphasized  that positron-electron annihilation with zero orbital angular momentum produces two photons 
described by the {\it entangled} wave function, Eq. (\ref{eq:wavefunc}), and therefore provides a particular entangled system for studying the EPR paradox. The authors calculated ratio $R$ and the angular distributions of Compton scattered photons for several types of  quantum states. They  confirmed $R\approx 2.85$ for the state Eq. (\ref{eq:wavefunc}) and concluded that $R=1$ for mixed separable state of annihilation photons described by the density matrix:

\begin{equation}
 \rho_\perp = \frac{1}{2}( \ket{{H_1}{V_2}}\bra{{H_1}{V_2}} + \ket{{V_1}{H_2}}\bra{{V_1}{H_2}}).
 \label{eq:density_matrix}
\end{equation}

According to the author's statement, the measured ratio  $R=2.0$ \cite{Wu} is experimental evidence of distant correlations leading to the EPR-paradox. It is worth noting that the experimental verification of entanglement was first discussed  seven years before the appearance of famous Bell's theorem \cite{Bell, Bell2}, which is basic instrument of modern experimental tests of entanglement. 

In subsequent years a series of experiments  \cite{Langhoff, Faraci, Kasday, Bertolini1981, Wilson1976, Bruno77} 
was performed to measure the ratio $R$ with better accuracy. The most precise results were obtained by Langhoff \cite{ Langhoff} ($R=2.47\pm0.07$) and by Kasday et al. \cite{Kasday} ($R=2.33\pm0.10$). 
The geometrical corrections that account for the finite solid angles of detectors provided the results consistent with the theoretically predicted value $R=2.85$. 

Quite recently Caradonna et al. \cite{Caradonna} derived the cross-sections for Compton scattering of annihilation photons in several (entangled and mixed separable) states using the matrix representation of the Klein-Nishina formula. Their results confirmed the credibility of Bohm-Aharonov finding and, consequently, the validity of relevant experimental tests. 

Strong polarization correlations for annihilation photons in entangled states and the theoretically predicted absence of correlations for separable states motivated the development a new generation of Positron Emission Tomography (PET), the so-called quantum-entangled QE-PET with Compton scattering reconstruction \cite{McNamara},\cite{Toghyani}. Attention is paid to the kinematic restrictions on the angular distributions of scattered gammas in attempts to suppress the scatter and random backgrounds and improve the quality of PET images. In recent years, significant efforts have been made to create working prototypes of PET using the reconstruction of the Compton scattering kinematics.  More recently, the research Jagiellonian Positron Emission Tomograph (J-PET) \cite{Moskal,Moskal_2019} was built of plastic scintillators.  Identification of double Compton scattering of photons in J-PET plastic bars makes it possible to determine the linear polarization of primary photons and to measure the angular correlations.

The verification of entanglement mentioned above relies exclusively  on the angular correlations of scattered annihilation photons. One might wonder why Bell's theorem \cite{Bell, Bell2},  or the Clauser- Horne-Shimony-Holt (CHSH) inequality \cite{CHSH} as a particular practical case, has not been applied here. The explanation is quite simple and is related to the low analyzing power of Compton polarimeters, which are the only tool for measuring the polarization of high-energy gamma rays. The analyzing power is strongly dependent on the energy of gamma-ray and is less than 0.7 for annihilation photons with an energy of 511 keV, even for an ideal polarimeter. This issue has been discussed at length by Clauser and Shimoni \cite{Clauser}, who pointed out that nonlocality in positron-electron annihilation can only be demonstrated if the analyzing power is greater than 0.83. In this case, the $S$-function in the CHSH inequality, reduced to the product of the analyzing powers of two Compton polarimeters, would exceed the required limit of 2. However, in reality, for annihilation photons, this  S-function can only reach a maximum of 1.4 even for an ideal polarimeter. 

Nevertheless, the $S$-function was measured in 1996 by Osuch et al. \cite{Osuch}. They counted the number of scattered photons in Compton polarimeters located at different azimuthal angles. The correlation coefficients were obtained from the coincidence of the count rates between two sets of polarimeters installed on opposite sides of the annihilation photon source. The $S$-function constructed from these coefficients perfectly reproduces the theoretically predicted behavior: 
\begin{equation}
 S = - p_0 (3\cos(2\phi) - \cos(6\phi)),
 \label{eq:Sfunction}
\end{equation}
where $\phi$ is the azimuthal angle between the polarimeters, and $p_0$ is the product of the analyzing powers of two opposite Compton polarimeters. Of course, the resulting maximum value of $S$ was well below the theoretical limit of $2$. Nevertheless, subsequent corrections by the measured values of the analyzing power of the polarimeters allowed the authors to state that the CHSH inequality is violated by twelve standard deviations.

The agreement of numerous experimental results with the predictions of quantum theory gives the impression of a complete understanding of the behaviour of a system of two annihilation photons and confidence in the entanglement of their initial state \cite{Caradonna}. However, the situation became rather uncertain in 2019 with the appearance of the theoretical work of Hiesmayr and Moskal \cite{Hiesmayr}, who applied an open quantum formalism to the Klein-Nishin formula and obtained the same Compton scattering cross section for both entangled and mixed separable states of annihilation photons. Their results fundamentally contradict previous theoretical considerations and, therefore, claim the incompleteness of existing experimental studies, which are based on the assumed difference in the Compton scattering of entangled and separable photon pairs. Moreover, as we will show below, the identical Compton scattering cross sections lead to the same $S$-functions in the CHSH inequality for entangled and separable mixed states. This raises the question of the relevance of Bell's theorem for testing quantum entanglement.

The results in ref.\cite{Hiesmayr} obviously affect the very possibility of using quantum entanglement in the development of PET imaging. To resolve apparent theoretical contradictions, Watts et al. \cite{watts} built a prototype PET with modern semiconductor gamma detectors and a passive Compton scatterer for one of the annihilation photons. They measured the angular correlation of scattered photons for two types of events: entangled photons without a passive scatterer and pre-scattered events with a passive scatterer placed in the path of one of the annihilation photons. In the latter case, the decohering process in the passive scatterer  leads to a separable state with lost entanglement. Unfortunately, the low sensitivity to the measured polarization and large statistical measurement errors did not allow to draw an unambiguous conclusion. New dedicated measurements with intact and prescattered  initial states of annihilation photons are required to solve the theoretical puzzle. For this purpose, we have developed a setup \cite{Abdurashitov} of Compton polarimeters with a large solid angle and an active intermediate Compton scatterer, which has been operating for about a year and a half at INR RAS.  In this paper, we report our rather unexpected findings on polarization correlations in Compton scattering of entangled and decoherent pairs of annihilation photons.

\section*{Methods}

Each annihilation photon has  an energy equal to the electron mass (511 keV), which is five orders of magnitude higher than the energy of optical photons. Such a sharp difference has both advantages and disadvantages in measuring polarization states. The need to use Compton polarimeters with a relatively low analyzing power has already been discussed and makes it problematic to use the CHSH inequality to test photon entanglement. At the same time, the high energy of photons makes it possible to create a separable, similar to that described by the Eq.\ref{eq:density_matrix}, mixed state through the Compton scattering of one of the photons. The selection of events with a sufficiently low energy release in a Compton scatterer would ensure the decohering process without noticeable distortion of the initial two-photon kinematics. Thus, annihilation photons provide a unique opportunity for direct comparison of both initially entangled and prepared separable mixed states. 

\begin{figure}[ht]
\centering 
\includegraphics[width=.5\textwidth]{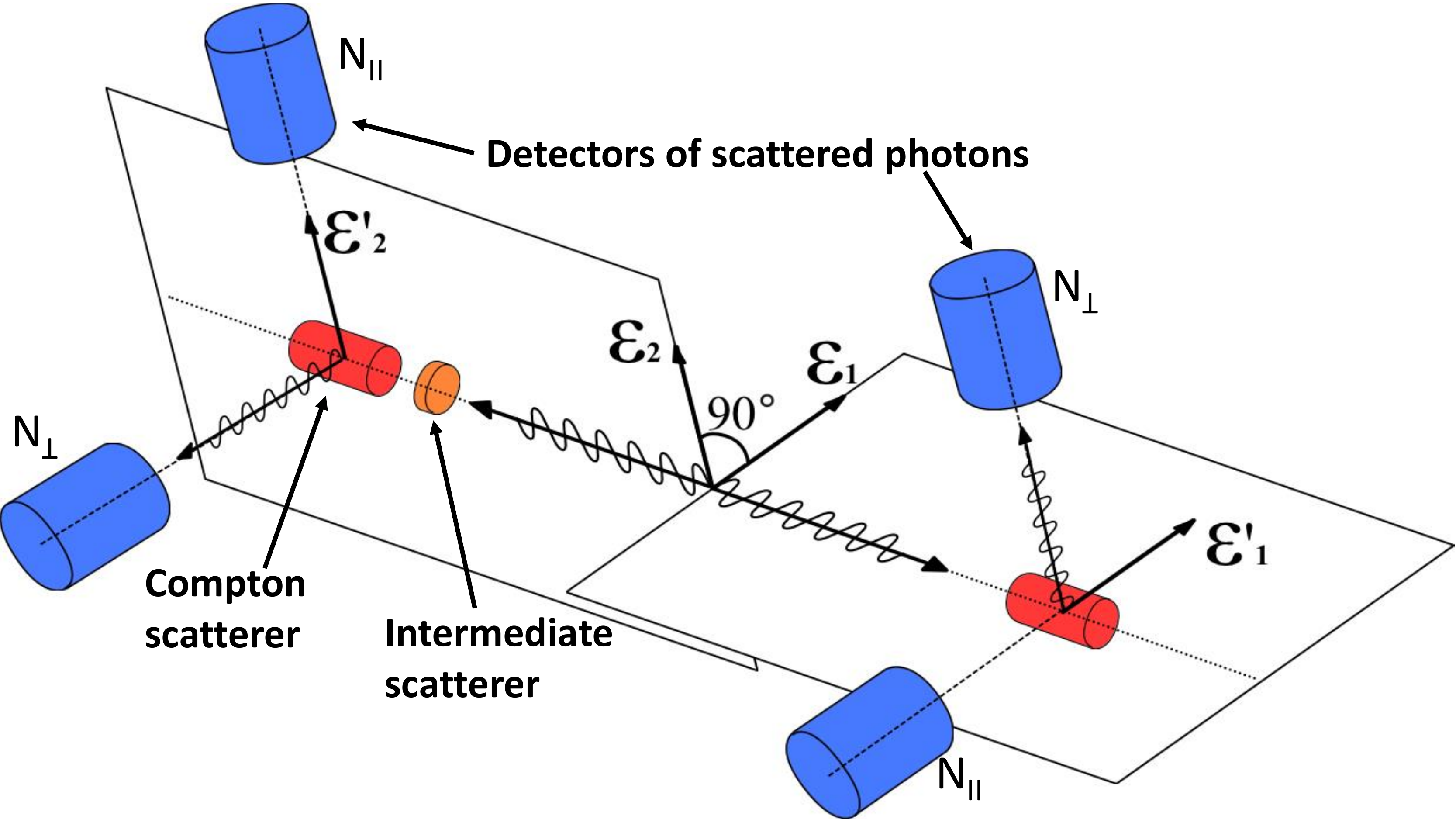}
\caption{\label{fig:i} Principal scheme for measuring polarization correlations of initially entangled and prepared decoherent states of annihilation photons. It includes two Compton polarimeters and an intermediate scatterer. Each Compton polarimeter consists of a scatterer and two orthogonal detectors of scattered gammas. $\varepsilon_1$ ($\varepsilon_2$) and $\varepsilon_1'$ ($\varepsilon_2'$) are the polarization vectors of the initial and scattered gammas, respectively. $N_{\parallel }$ and $N_{\perp}$ denote detectors parallel (perpendicular) to the initial polarization vector.}
\label{scheme_polarimeter}
\end{figure}

The principles for measuring the polarization correlations of these two types of quantum states are illustrated in Fig.~\ref{scheme_polarimeter}. 
Two Compton polarimeters are required for measuring both photons with opposite momenta. Each polarimeter consists of a Compton scatterer and two detectors of scattered gammas arranged orthogonally.  An intermediate scatterer is placed in the path of one of the initial photons to create a tagged subset of pre-scattered events. Due to the decohering process during the interaction, entanglement is lost for the corresponding pairs of tagged photons.

The measurements are based on the dependence of the differential cross section of Compton scattering on the polarization direction, which is given by the well-known Klein-Nishina formula \cite{Nishina}:
\begin{equation}
\frac{d\sigma}{d\Omega}=\frac{ (r_e\epsilon)^2}{2}\cdot \left ( \epsilon+\frac{1}{\epsilon}-2\sin^2{\theta}\cdot\cos^2{\phi} \right),
\label{eq:kl-ni}
\end{equation}
where ${r_e}$ is the classical electron radius, $\epsilon \equiv E_1/E$, $E$ ($E_1$) is the energy of the incident (scattered) photon, $\theta$ is the scattering angle and $\phi$ is the angle between the scattering plane and the direction of polarization of the incident photon. The scattered photon energy is
${E_1=E\cdot m_e/(m_e+E\cdot(1-\cos{\theta})})$, 
where $m_e$ is the electron mass. As follows from Eq.\ref{eq:kl-ni}, photons scatter predominantly orthogonally to the polarization plane. 

Like optical polarimeters, the Compton polarimeter has two main components. Namely, the Compton scatterer measures the polarization of photons in place of a conventional polaroid, and the detectors of scattered photons play the role of photodetectors.

The main characteristic of a Compton polarimeter, which determines the sensitivity to the measured polarization, is the analyzing power:
$A(\theta)=\frac{N_\perp-N_\parallel}{N_\perp+N_\parallel}$, 
where ${N_\perp}$ $({N_\parallel})$ denotes the number of registered events in counters located perpendicular (parallel) to the polarization of the incident photons.  Equivalently, at large statistics,

\begin{equation}
A(\theta)=\frac{\frac{d\sigma} {d\Omega}(\theta,\phi=90^\circ)-\frac{d\sigma}{d\Omega}(\theta,\phi=0^\circ)}{\frac{d\sigma} {d\Omega}(\theta,\phi=90^\circ)+\frac{d\sigma}{d\Omega}(\theta,\phi=0^\circ)}. 
\label{eq:apow1}
\end{equation}

Using Eq. \ref{eq:kl-ni}, the analyzing power is obtained as \cite{Knights}:
\begin{equation}
A(\theta)=\frac {\sin^2{\theta}}{E_{1}/E+E/E_{1}-\sin^2{\theta}} .
\label{eq:apow}
\end{equation}

As follows from Eq. \ref{eq:apow}, for a given energy of the initial gamma, the analyzing power depends on the scattering angle.
For completely polarized photons with an energy of 511 keV, the analyzing power reaches a maximum of $A=0.69$ at $\theta=82^\circ$.

In the case of an entangled pair of photons with mutually orthogonal polarizations, the probability of Compton scattering at scattering angles $\theta_1, \theta_2$ is given by the following expression \cite{Snyder},\cite{Caradonna}, \cite{Hiesmayr} :
\begin{equation}
P(\Delta\phi)={k_1 k_2  \left(1-\alpha(\theta_1)\alpha(\theta_2)\cos(2\Delta\phi)\right)},
\label{eq:prob}
\end{equation}
where $k_1$, $k_2$ are the kinematic factors for the first and second scattered photons, $ \Delta\phi$ is the angle between scattering planes, the parameters $\alpha(\theta_1)$ and $\alpha(\theta_2)$ coincide with the analyzing power of the corresponding Compton polarimeter. The product of the analyzing powers is equal to the modulation factor $\mu$ \cite{kozuljevich} which determines the sensitivity of a setup to the measured polarization. Therefore, Eq.\ref{eq:prob}, together with the dependence of photon counts on the azimuthal angle, also gives the modulation factor $\mu$ of the experimental setup.

\section*{Experimental setup}

\begin{figure}[ht]
\centering 
\includegraphics[width=.5\textwidth]{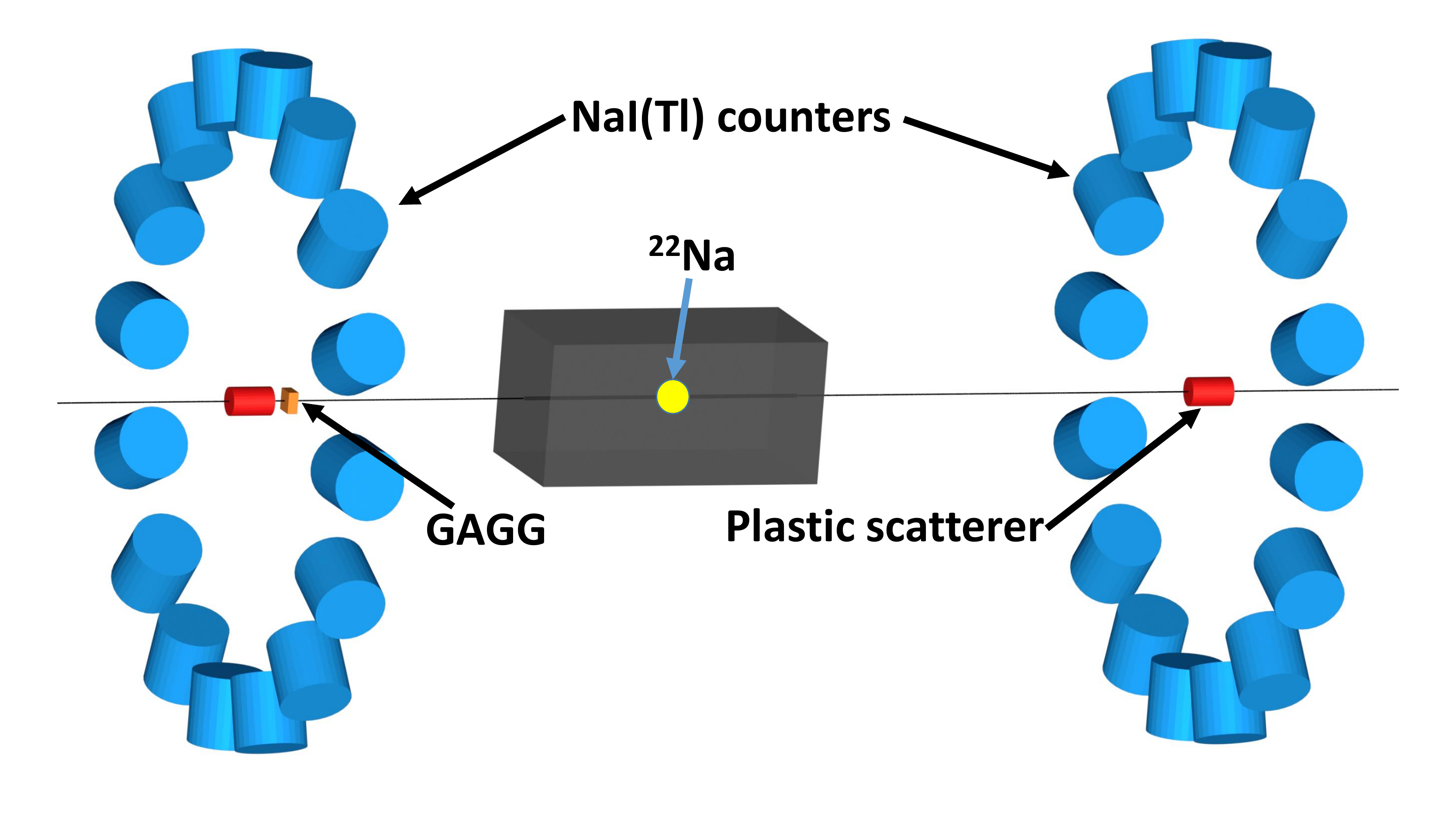}
\caption{\label{fig:scheme_setup} Scheme of a two-arm experimental setup. Each arm consists of a plastic scatterer on the setup axis and 16 NaI(Tl) counters orthogonal to the axis. The $^{22}$Na source of positrons is placed in a lead collimator between the arms closer to intermediate GAGG scatterer for preparing annihilation photons in  decoherent states. }
\label{fig:setup_scheme}
\end{figure}

Based on the above principles an experimental setup \cite{Abdurashitov} was constructed to measure the polarization correlation of annihilation photons in both entangled and separable states. The setup comprises two equivalent arms of Compton polarimeters and a $^{22}$Na  positron  source placed between these arms, as shown in Fig.\ref{fig:setup_scheme}. 

A source of positrons with an activity of $\sim50$ MBq was fabricated by irradiating a 1 mm thick aluminum plate with 130 MeV protons at the INR RAS isotope facility \cite{Zhuikov}. The source is located in a horizontal hole in a lead cube  providing collimated beams of annihilation photons in opposite directions from the source. Each arm of the setup consists of a plastic scintillation scatterer and a ring of 16 NaI(Tl) scintillation detectors of scattered photons with an azimuthal angle between adjacent detectors of $22.5^{\circ}$. NaI(Tl) counters detect photons scattered at an angle of about $90^{\circ}$.  Each pair of orthogonal NaI(Tl) counters and a plastic scatterer of the same arm form an elementary Compton polarimeter. Due to the azimuthal symmetry of the setup, each NaI(Tl) counter registers the vertical or horizontal polarization component   depending on the orientation of the scattering plane.  The chosen layout leads to a compensation for possible systematic errors caused by different efficiencies and inaccuracies in the positions of the NaI(Tl) counters. 
 
The distance between the plastic scatterers is about 70 cm. To break down the entanglement of annihilation photons in a subset of events, an intermediate thin scatterer, a gadolinium-aluminum-gallium garnet (GAGG) scintillator, is placed next to one of the plastic scatterers. The $^{22}$Na positron source is located 10 cm closer to the arm with the GAGG scintillator to ensure that the first photon interaction occurs in the intermediate scatterer.
 
The GAGG scatterer is the key element that separates events in entangled or decoherent states.  An interaction in the GAGG scintillator means that a pair of initially entangled annihilation photons has undergone a decoherence process. Therefore, the reliability of interaction identification in the intermediate scatterer is the most important feature of the setup.

\begin{figure*}[htbp]
\centering 
\includegraphics[width=.46\textwidth]{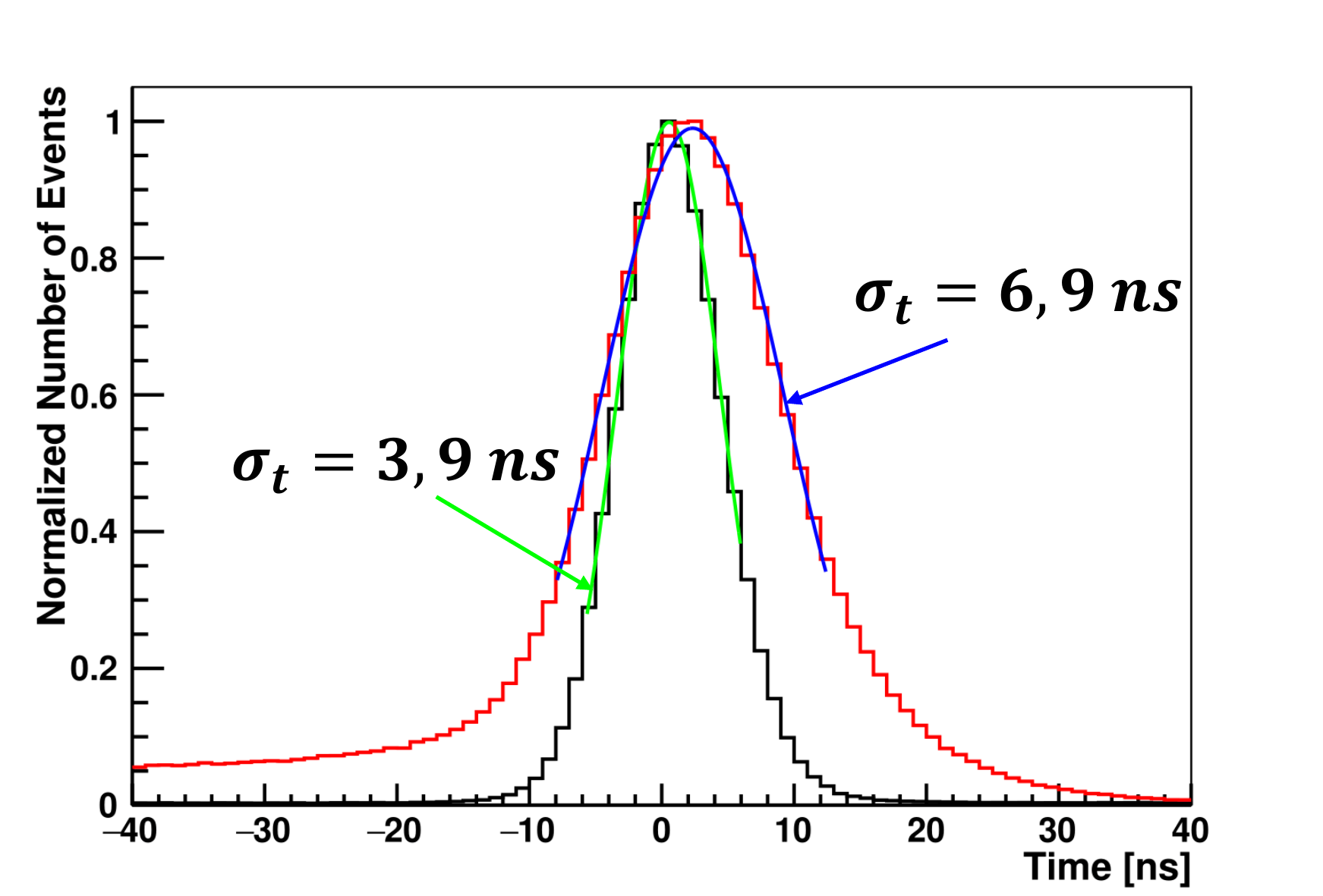}
\qquad
\includegraphics[width=.47\textwidth]{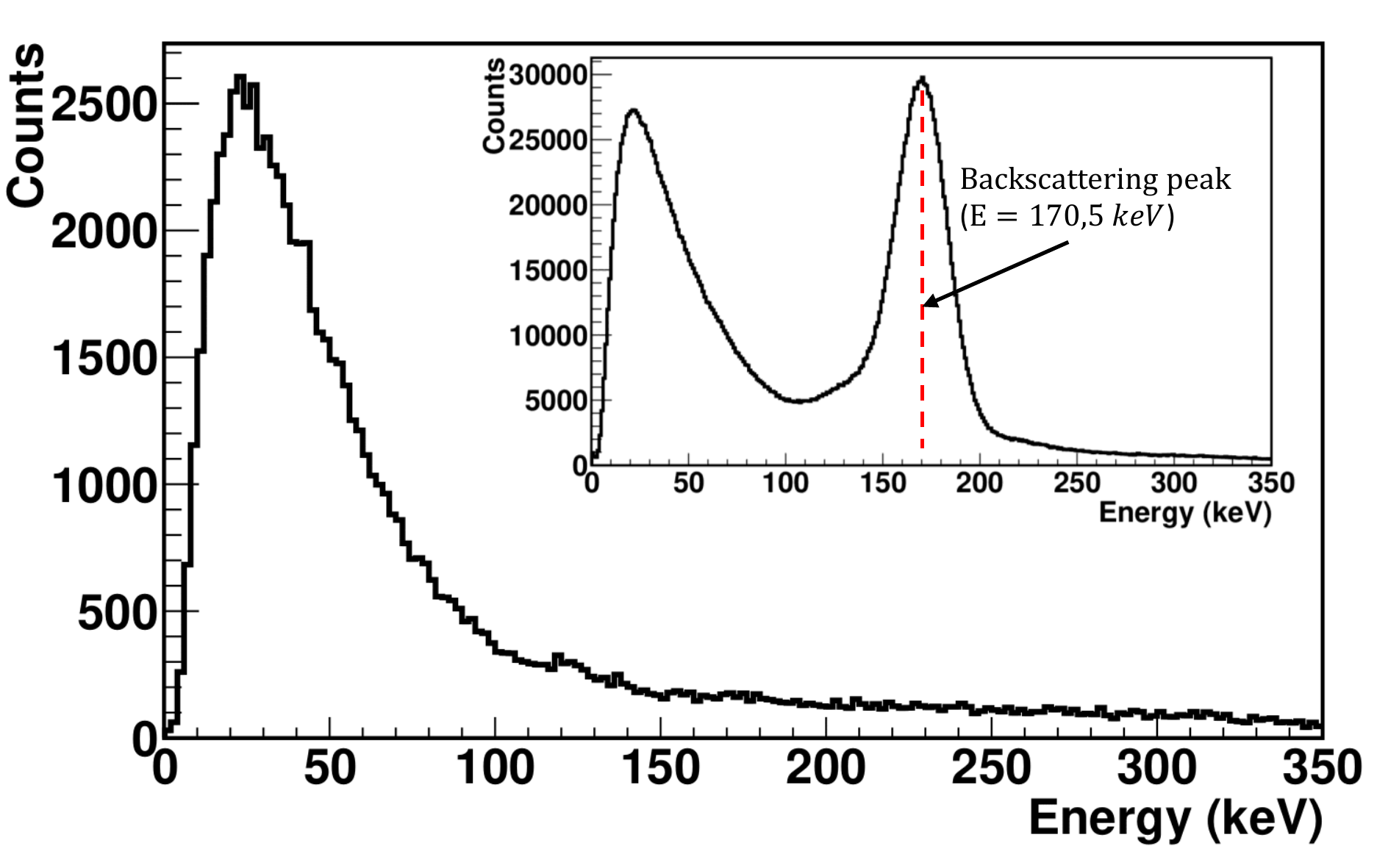}
\caption{\label{fig:i} Left - time coincidence spectra between signals in  intermediate GAGG and plastic scatterers. Red and black lines correspond to events with energy release in GAGG in the ranges of 2–40 keV and 40–120 keV, respectively.  The numbers indicate the time resolution for these two cases. Right - the energy spectra in GAGG scatterer for events within the true time coincidence peak. Here, events are selected that hit the NaI(Tl) counters. Insert shows the extended GAGG energy spectrum for all events, regardless of the hits in NaI(Tl) counters. The prominent peak at 170.5 keV corresponds to photons backscattered by the adjacent plastic scatterer and absorbed by GAGG. This peak is used for energy calibration of the intermediate scatterer.}
\label{fig:GAGG}
\end{figure*}

Energy deposition and signal timing are measured in all scintillators, NaI(Tl) detectors, intermediate GAGG and plastic scatterers. The timing and amplitude of the signals used to identify the interaction in the GAGG scintillator are shown in Fig.\ref{fig:GAGG}.

The time coincidence spectra of signals from intermediate and plastic scatterers are presented for two cases, with low and high energy deposition in the GAGG. In the first case, the scattering angles in GAGG are small and the influence on the momentum of the initial annihilation photon is minimal. In these events, the noise of the photodetector and readout electronics stronger affects the time resolution compared to events with higher energy deposition.

Fig.\ref{fig:GAGG}, on the right, shows the energy spectra in the intermediate scatterer for events in true time coincidence window with hits in the NaI(Tl) counters. The high light yield of GAGG \cite{GAGG1,GAGG2} makes it possible to detect recoil electrons with an energy threshold of several keV. The permanent energy calibration of GAGG is performed by detecting 170.5 keV energy deposition from photons backscattered by an adjacent plastic scatterer and absorbed in GAGG.

We define the tagging of events of various types analyzed in polarimeters as follows. A pair of initial photons is considered entangled if no interaction is observed in the intermediate GAGG scatterer. Otherwise, the detected energy in the GAGG scatterer means that the photons have undergone decoherence.

The sensitivity of the setup to polarization measurements was studied by the Monte Carlo simulation using the latest version of Geant4 \cite{Geant4, Geant2016} particle simulation framework, where the theoretical formula Eq.\ref{eq:prob} is implemented for initial annihilation photons. Since the NaI(Tl) detectors are located orthogonally to the Compton scatterers, the  theoretical ratio of counts for perpendicular and parallel orientations of the scattering planes can be $R=2.6$ which is less than the maximum $R=2.85$ for the optimal $82^\circ$ scattering angle. In a setup with a  realistic detector geometry, photons are detected in the range of scattering angles of $80^\circ-100^\circ$, which leads to an additional reduction to $R=2.40$. The modulation factor for entangled photons equals to $\mu=0.41$, which is about 15\% less than the maximum theoretical value $\mu=0.48$ for the optimal scattering geometry.

Finally, Monte Carlo simulations estimated the contribution of systematic errors caused by possible inaccuracies in the positions of the detectors and the positron source. Due to the azimuthal location of the detectors, systematic errors turned out to be almost an order of magnitude smaller than the statistical errors achieved and were not taken into account.

\section*{Results}

\subsection*{Experimental spectra}

\begin{figure}[hbtp]
\centering 
\includegraphics[width=.465\textwidth]{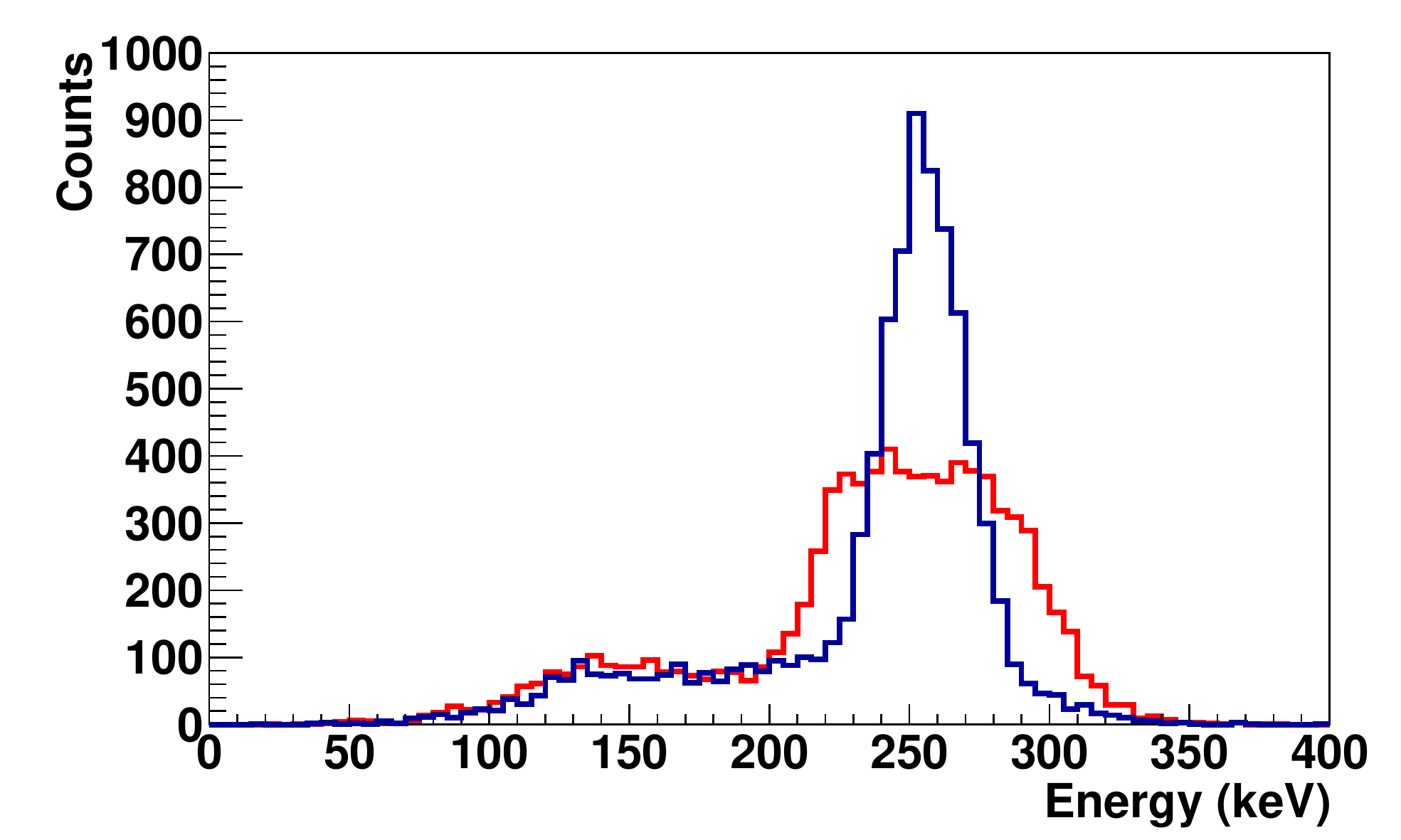}
\includegraphics[width=.465\textwidth]{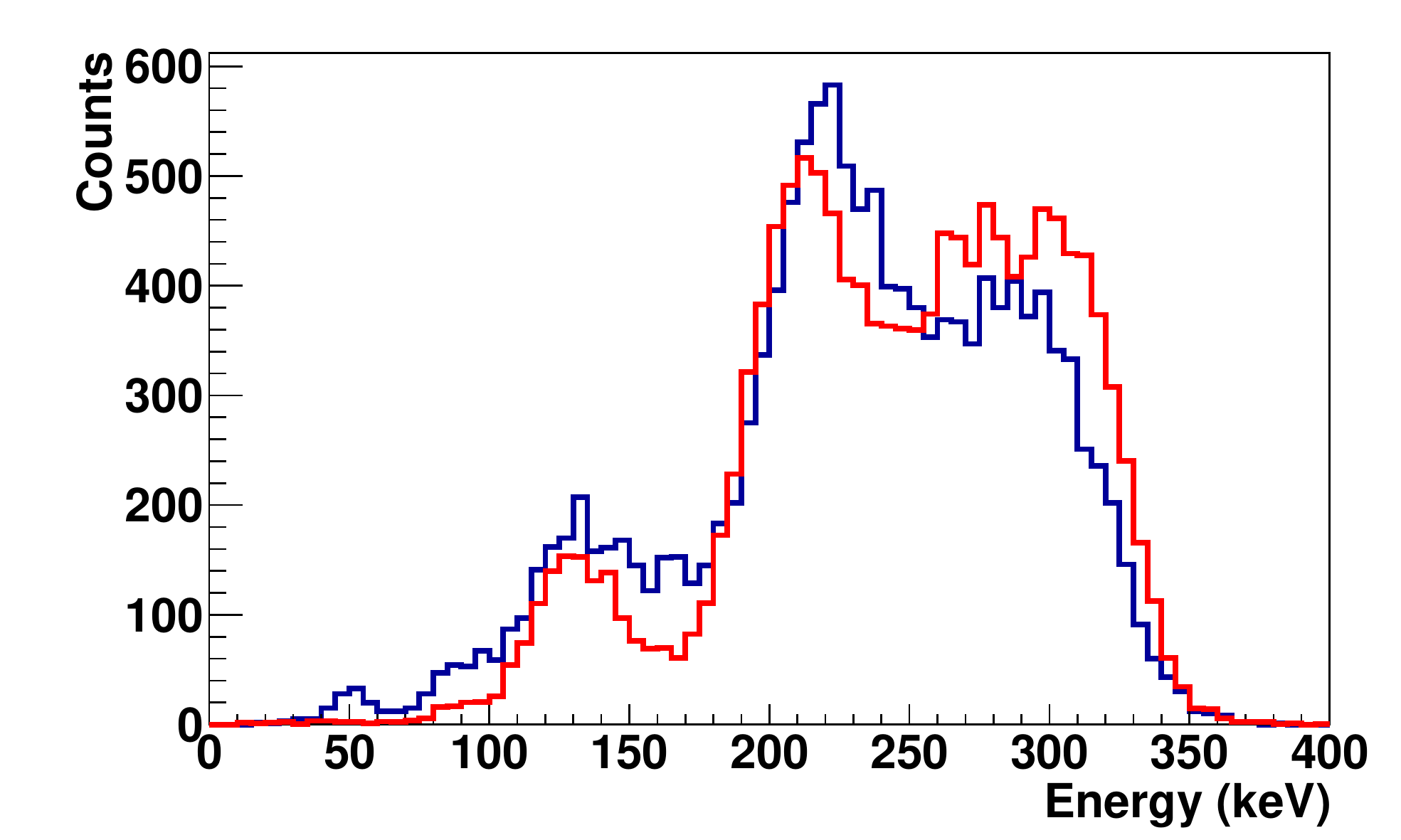}
\qquad

\caption{Left - energy spectra in NaI(Tl) counters for events in entangled states without energy deposition in the intermediate GAGG scatterer (blue line) and in decoherent states with energy deposition in the intermediate GAGG scatterer below 30 keV (red line).
Right - the energy spectra in NaI(Tl) counters for decoherent photons with energy deposition in the intermediate GAGG scatterer between 30 keV and 110 keV.  The blue line is the experimental data, the red line is the result of the Monte Carlo simulation.
} 
\label{spectra}
\end{figure}

Identification of the  Compton scattering kinematics is carried out by monitoring the energy release in all active elements of Compton polarimeters, namely, in NaI(Tl) detectors of scattered photons, plastic scatterers, and the intermediate GAGG scatterer. Scattering of the initial annihilation photons at $90^{\circ}$ releases on average an equivalent energy of 255 keV both in the plastic scatterer and in the NaI(Tl) counter.  According to Monte Carlo simulations, in a setup with a  realistic detector geometry, photons are detected in the range of scattering angles of $80^\circ-100^\circ$ with energy deposition in NaI(Tl) from 235 keV to 280 keV. This is illustrated in Fig.\ref{spectra}, which shows energy deposition spectra in NaI(Tl) detectors for various Compton scattering kinematics.

The relatively narrow NaI(Tl) energy peak for entangled states reflects the range of scattering angles of registered photons.  The situation is completely different for  decoherent states after the interaction of photons with an intermediate scatterer.  Even a few percent energy loss (below 30 keV) of the original 511 keV photons in the GAGG scintillator leads to a significant distortion of the energy spectrum. This is the effect of double Compton scattering of the initial annihilation photon in both GAGG and plastic scintillators. After the first scattering in GAGG, the photon deviates within a few degrees from its original direction. Subsequent interaction in a plastic scintillator increases the ranges of scattering angles and energy deposition of photons registered in NaI(Tl).

Events with a higher energy deposition in the intermediate scatterer (up to 110 keV) form a complex structure in the NaI(Tl) spectrum.  Visible peaks in this energy spectrum correspond to particular cases of the kinematics of Compton scattering. The Monte Carlo simulation of the double Compton scattering of 511 keV photons confirmed that the observed structure in the NaI(Tl) spectrum reflects the kinematics of the first scattering in the GAGG scintillator, which leads to distinct groups of events. 
These particular cases of Compton scattering and the correlation between energy depositions in the NaI(Tl) and GAGG scintillators are shown in Fig. \ref{scatter_plot}.

\begin{figure*}[htbp]
\centering 
\includegraphics[width=.99\textwidth]{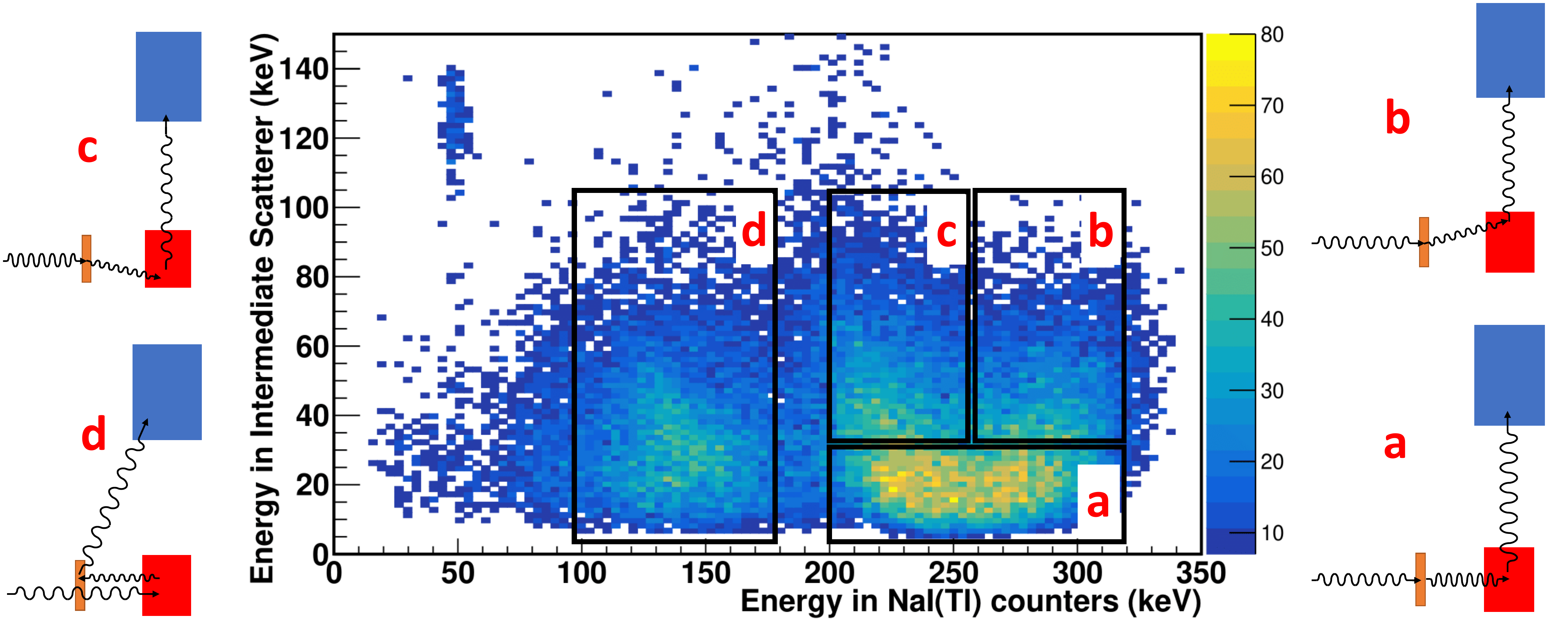}
\caption{\label{scatter_plot} Correlation between energy depositions in the intermediate GAGG scatterer and in NaI(Tl) counters. Several groups of events in black boxes marked as "a"," b", "c" and "d" can be distinguished. They correspond to different Compton scattering kinematics in the GAGG scintillator, shown in the diagrams to the left/right of the correlation plot. The brown, red and blue boxes in the diagrams mark the intermediate GAGG scatterer, the plastic scatterer and the NaI(Tl) counter, respectively. }
\label{scatter_plot}
\end{figure*}

In this figure,  several distinct groups of events can be identified, labeled as classes "a", "b", "c", and "d". The double Compton scattering kinematics for these events can be determined using a simple analytical calculation or Monte Carlo simulation.
Group "a" represents scattering processes with an electron recoil energy in the GAGG scintillator below 30 keV and, accordingly, with the smallest scattering angles.  

The range of scattering angles by the GAGG in the  "b" and "c"  regions is identical and it is larger as compared to the "a"-events.  However,  events in the "b” region have maximum energy deposition in NaI(Tl)  detectors from two sequential Compton scatterings  with the first scattering in the GAGG in the direction of the NaI(Tl)  counter.
In contrast to these events, the "c" group  represents a double Compton scattering with the sum of scattering angles exceeding $90^\circ$, since the first scattering in the GAGG scintillator deflects the photon in the opposite direction from the NaI(Tl) counter.  Therefore, more energy is left in the plastic scintillator and less for the NaI(Tl) detector remains.

The most complex kinematics of double Compton scattering is observed for the ``d" group of events with the lowest energy registered by NaI(Tl). The original annihilation photons in these events pass through the GAGG scintillator without interaction and are backscattered in the plastic scatterer. The second scattering at about $90^\circ$ occurs in the GAGG scintillator, followed by photon registration in NaI(Tl) counters. Unlike other classes of events with relatively small scattering angles in GAGG, the "d" group represents Compton scattering at maximum angles  about $180^\circ$  and with maximum energy losses in plastic scatterer.

\subsection*{Angular distributions of scattered photons.}
Previous experiments with entangled annihilation photons have mainly measured the dependence of the number of counts of scattered photon detectors on the azimuthal angle between these detectors. This dependence is described by Eq.\ref{eq:prob} and has a cosine-like behaviour.  As follows from the formula, the ratio of counts $R$ for perpendicular and parallel orientations of the scattering planes reaches a maximum of $R=2.85$ if both photons are scattered at an angle of $82^0$ to their initial momenta. According to Eq.~\ref{eq:prob}, the number of counts in detectors of scattered photons can be approximated as:
\begin{equation}
N(\phi)=A-B\cos(2\Delta\phi),
\label{coincidences}
\end{equation}
where $\Delta\phi$ is the angle between scattering planes, A and B are the fit constants. As follows from Eq.~\ref{eq:prob} and Eq.~\ref{coincidences}, the ratio $R$ equals to $R=({A+B})/({A-B})$, while the modulation factor is $\mu=B/A$. 

\begin{figure}[ht]
\centering 
\includegraphics[width=.435\textwidth]{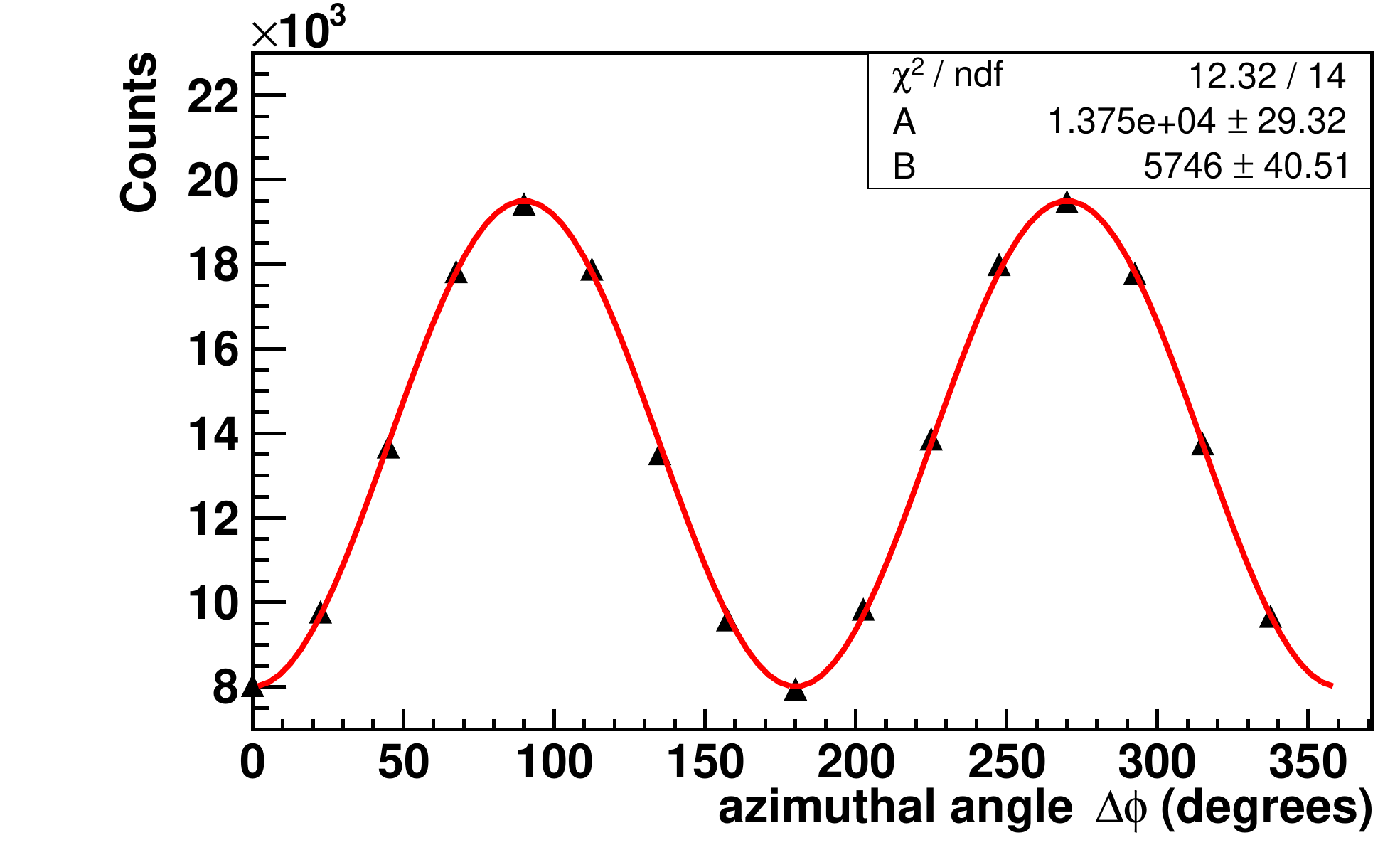}
\qquad
\includegraphics[width=.435\textwidth]{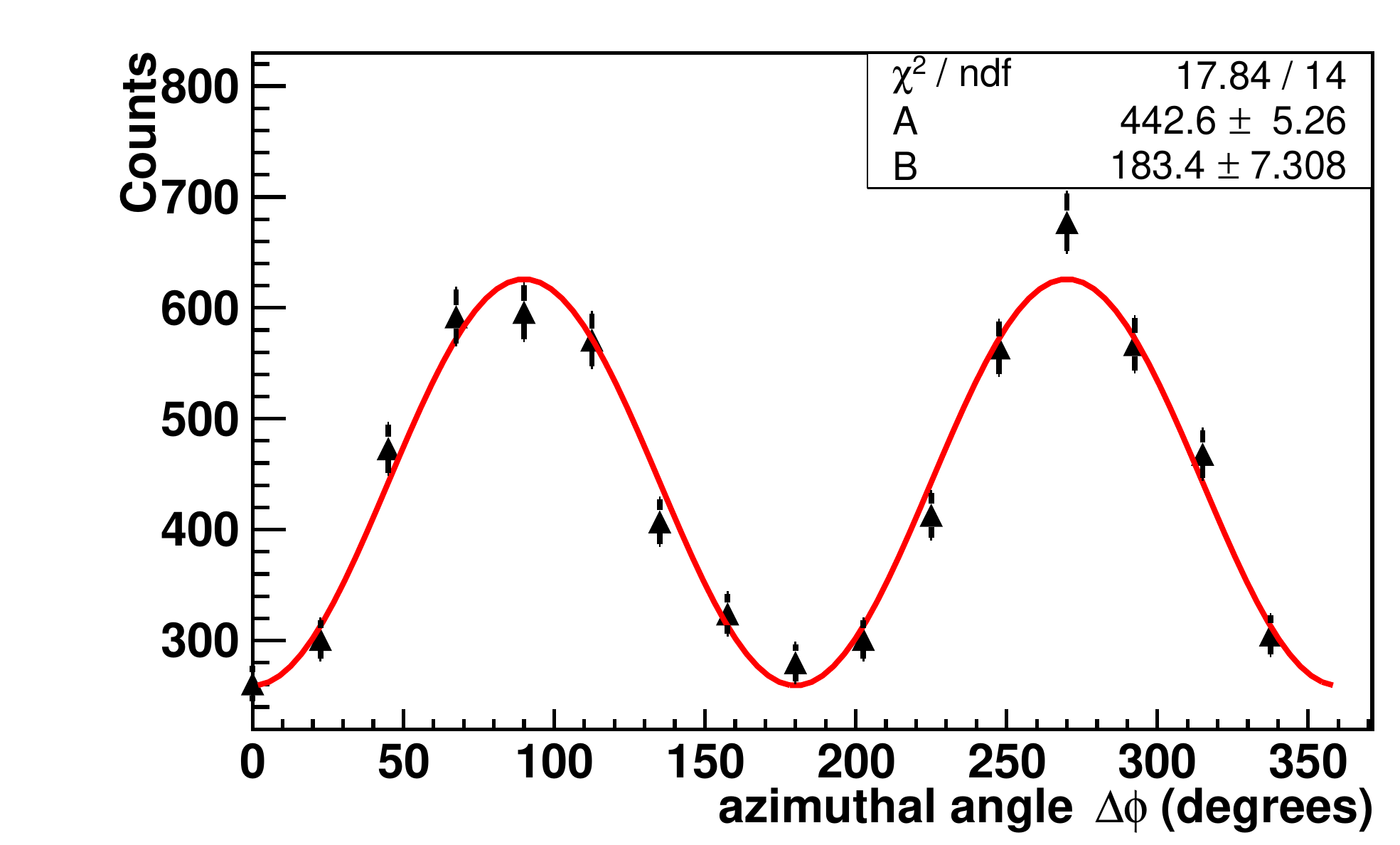}
\caption{\label{fig:i} Dependence of coincidence counts in NaI(Tl) detectors on the azimuthal angle between these detectors for entangled (left) and decoherent (right) initial states of photon pairs. The solid line corresponds to the fitting function Eq.~\ref{coincidences}.}
\label{all_events}
\end{figure}

 In our experiment, the  data set without energy deposition in intermediate GAGG scatterer is associated with entangled photons, Eq.~\ref{eq:wavefunc}.  In this case, the modulation factor coincides with the product of the analyzing powers of the corresponding Compton polarimeters, $\mu=\alpha({\theta_1})\cdot \alpha({\theta_2})$, see Eq. \ref{eq:prob}. Other events with  interaction in the GAGG scatterer correspond to  decoherent photon pairs in the initial state.   
Experimental angular distributions for these two types of events  with different initial quantum states are shown in Fig.~\ref{all_events}. The data were approximated by the function Eq.~\ref{coincidences}.  For the entangled photons, the fitting parameters give  $R = 2.435 \pm 0.018$, which is consistent  with the Geant4 Monte Carlo simulation, taking into account that the NaI(Tl) detectors are set orthogonal to the symmetry axis and register photons scattered in range $80^\circ - 100^\circ$. The modulation factor  equals to $\mu=A^2 = 0.418 \pm 0.003$, which is about 15\% less than the maximum theoretical value of 0.48 for ideal Compton polarimeters.

The second plot in Fig.~\ref{all_events} shows the angular distributions for the entire set (classes "a", "b", "c" in Fig.~\ref{scatter_plot}) of decoherent events, excluding backscattered ones. Surprisingly, decoherent events exhibit almost the same azimuthal behavior with a ratio $R = 2.41 \pm 0.10$ and a modulation factor $\mu = 0.414 \pm 0.017$. 

Interaction in GAGG is equivalent to quantum mechanical measurement. After the measurement, the participating photon must be in the ``H" or ``V" state with equal probability. The other photon must collapse into a state of opposite polarization if there is no exchange of angular momentum with the electron, which is true at low energies $E \ll m_e$ or forward scattering $\theta^2  \ll m_e/E$, Eqs. (87.20-21) in ref.\cite{berestetskii}. Therefore, after scattering in GAGG, under one of these conditions, decoherent photons must  be in a mixed separable state Eq.~\ref{eq:density_matrix}. Events of classes "a", "b", "c" satisfy the forward scattering condition. Then the observed cosine-like behavior contradicts earlier theoretical findings~\cite{Bohm}, see also recent ref.~\cite{Caradonna}.  On the contrary, our experimental results are in agreement with  the recent prediction~\cite{Hiesmayr}  of the equivalent Compton scattering cross section for photons in entangled and mixed separable states, Eq.~\ref{eq:wavefunc} and Eq.~\ref{eq:density_matrix} respectively.

\begin{figure*}[htbp]
\centering 
\includegraphics[width=1\textwidth]{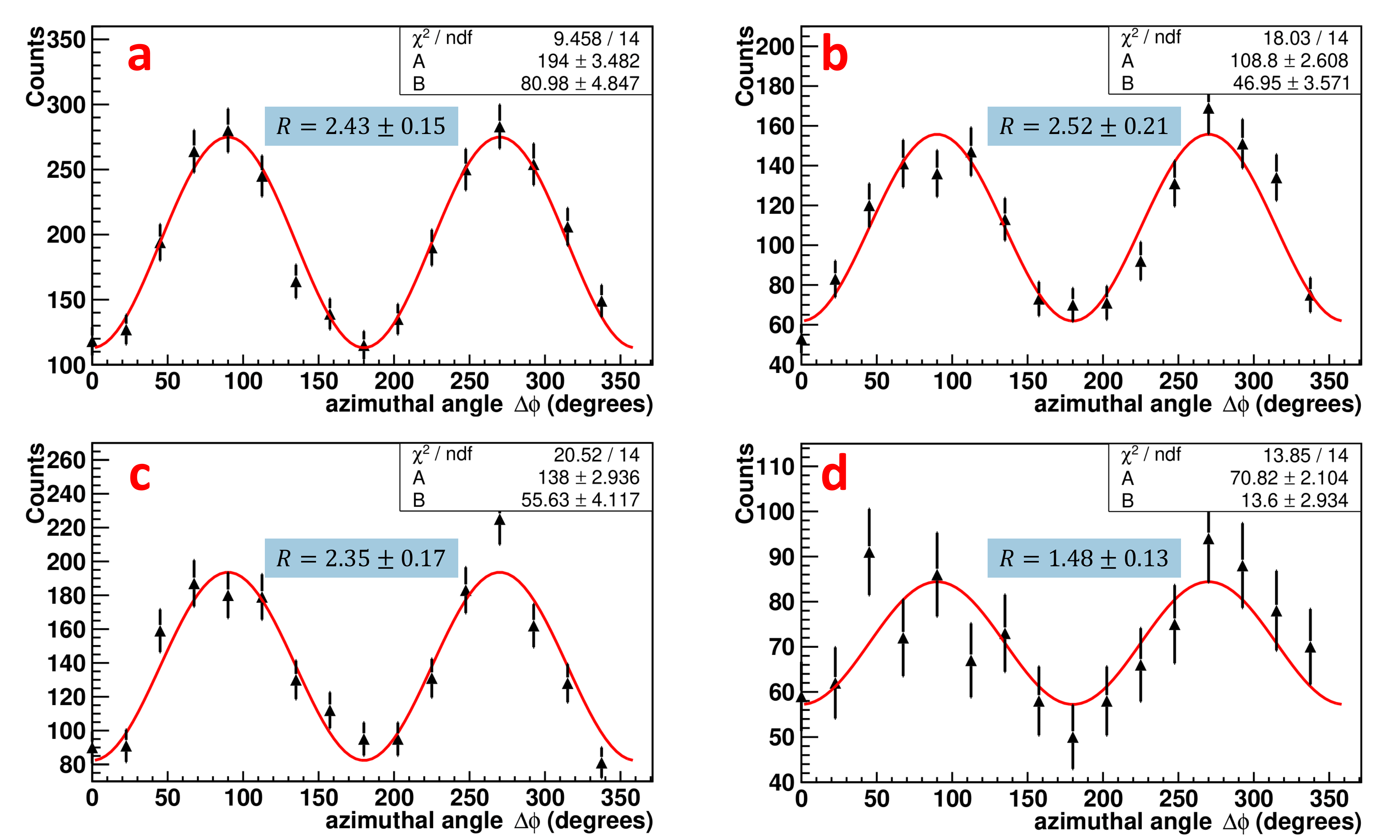}

\caption{\label{fig:i} Dependence of coincidence counts in NaI(Tl) detectors on the azimuthal angle between these detectors for four classes ( "a", "b", "c" and "d" in Fig.~\ref{scatter_plot}) of decoherent events. The solid line corresponds to the fitting function Eq.~\ref{coincidences}. The numbers in the blue area on the graphs indicate the $R$ ratio for the corresponding class of events.}
\label{four_groups}
\end{figure*}

 To get more information and insight, angular distributions were constructed for all classes of decoherent events classified in Fig.~\ref{scatter_plot}.  The dependencies of the number of scattered photons registered in the NaI(Tl) counters on the azimuthal angle between these photons are shown in Fig.~\ref{four_groups}.  From the fitting parameters presented in these plots, one can calculate the ratio $R$ of the numbers of counts for the perpendicular and parallel orientations of the scattering planes, which are very close to each other, $R_a = 2.43 \pm 0.15$, $R_b = 2.52 \pm 0.21$, $R_c = 2.35 \pm 0.17$ for classes "a"," b", "c", respectively. 

The smallest ratio $R_d = 1.48 \pm 0.13$ is observed for class "d" events. This is due to the partial depolarization of backscattered photons, as follows from the analysis of the Klein-Nishina formula for linearly polarized gamma rays \cite{berestetskii}. Namely, according to Eq.~(87.12) in ref.~\cite{berestetskii}, with $E = m_e$ and backscattering by $180^{\circ}$, one fifth of photons change from vertical to horizontal polarization and vice versa. Note, that in \cite{Hiesmayr}  the possibility of such transitions is missed. Adding the corresponding fraction of
$$ \rho_\parallel =  \frac{1}{2}( \ket{{H_1}{H_2}}\bra{{H_1}{H_2}} + \ket{{V_1}{V_2}}\bra{{V_1}{V_2}})$$ 
to the density matrix Eq.~\ref{eq:density_matrix} we find that the modulation factor reduces by a factor of $0.6$. We should expect $R \approx 1.66$ for such a depolarized state, which is consistent with the experimental result, Fig.~\ref{four_groups}. Note, that $R = 1$ if $\rho =   \frac{1}{2}(\rho_\perp + \rho_\parallel)$. 
 
Since the angular distributions of scattered photons turned out to be the same for entangled and separable initial states, our results show that these dependences cannot be used to experimentally test the entanglement of annihilation photons. 

\subsection*{$S$-function in CHSH inequality.}

The CHSH inequality as a particular variant of Bell's theorem is used in  classical  experiments with pairs of entangled optical photons \cite{Aspect1,Aspect2,Aspect3}.   
A suitable setup includes two dual-channel  polarimeters separating  orthogonal linear polarization parallel and perpendicular to some arbitrary $\vec{a}$ and $\vec{b}$ directions for the first and second photons, respectively.  By measuring the coincidence rate $R$ of photons with different spin orientations, the correlation coefficients can be constructed as:
\begin{equation}
E(\vec{a},\vec{b})= 
\frac{R_{a_{\parallel}b_{\parallel}} + R_{a_{\perp}b_{\perp}}-R_{a_{\parallel}b_{\perp}}-R_{a_{\perp}b_{\parallel}}}{R_{a_{\parallel}b_{\parallel}} + R_{a_{\perp}b_{\perp}}+R_{a_{\parallel}b_{\perp}}+R_{a_{\perp}b_{\parallel}}},
\label{correlation_coefficient}
\end{equation}
where the indexes $a_{\parallel}$, $a_{\perp}$, $b_{\parallel}$ and $b_{\perp}$ denote the photons with parallel or perpendicular polarization for the direction $\vec{a}$ or $\vec{b}$, respectively. 

By measuring the correlation coefficients for four different polarimeter orientations, $\vec{a}, \vec{b}, \vec{a}^\prime, \vec{b}^\prime$, the following function can be composed:\begin{equation}
S=E(\vec{a},\vec{b}) - E(\vec{a},\vec{b}^\prime) + E(\vec{a}^\prime,\vec{b}) + E(\vec{a}^\prime,\vec{b}^\prime).
\label{S_function}
\end{equation}
In the case of entangled photons, $S$-function reaches a maximum of $\lvert S\rvert= 2\sqrt{2}=2.83$ for certain optimal polarimeter orientations:
\begin{gather*}
 (\vec{a},\vec{b})=(\vec{a}^\prime,\vec{b}^\prime)=(\vec{a}^\prime,\vec{b})=\phi_{opt}\\
(\vec{a},\vec{b}^\prime)=(\vec{a},\vec{b})+(\vec{a}^\prime,\vec{b}^\prime)+(\vec{a}^\prime,\vec{b}),
\end{gather*}

where   $(\vec{a},\vec{b})$ denotes an angle between vectors $\vec{a}$ and $\vec{b}$, and the optimal azimuthal angles $\phi_{opt}$ are multiples of $\phi=22.5^\circ$.
This clearly violates the CHSH inequality $\lvert S\rvert\leq2$ that follows from Bell's theorem if  underlying  hidden local variables would exist.  

As discussed by Clauser and Shimony \cite{Clauser}, the CHSH inequality is never violated for annihilation photons due to the low efficiency (analyzing power) of Compton polarimeters.  To map  the real-world experiments with non-ideal polarimeters to the ideal case, the $S-$function is usually normalized to the product of corresponding efficiencies.   Moreover, even without such a normalisation,  measurements of annihilation photons in entangled and  decoherent states on the same setup allow a direct comparison of the $S$-function for these two types of events. One should expect a fundamentally different behavior of this function for entangled and separable states, since in the latter case there are no nonlocal correlations.

Since  in our setup the angle between adjacent detectors of scattered photons is $22.5^\circ$,  the procedure  for measuring the correlation coefficients and the $S$-function is straightforward (see also Osuch et al. \cite{Osuch}). The elementary Compton polarimeter consists of a scatterer and two orthogonal detectors of scattered photons, as shown in Fig~\ref{scheme_polarimeter}. To measure the correlation coefficient $E(\vec{a},\vec{b})$, two elementary Compton polarimeters are requested on opposite sides of the setup with orientations $\vec{a}$ and $\vec{b}$. As an example, Fig.~\ref{scheme_polarimeter} shows a scheme for measuring the correlation coefficient $E(90^{\circ})$. The $S$-function was determined using four polarimeters with orientations $\vec{a}$, $\vec{a}^\prime$ on one side and $\vec{b}$, $\vec{b}^\prime$ on the other. There are 16 Compton polarimeters on each side of the setup. And the count rates for a given correlation coefficient are summed for all relevant combinations of polarimeters.

\begin{figure}[ht] 
\centering 
\includegraphics[width=.51\textwidth]{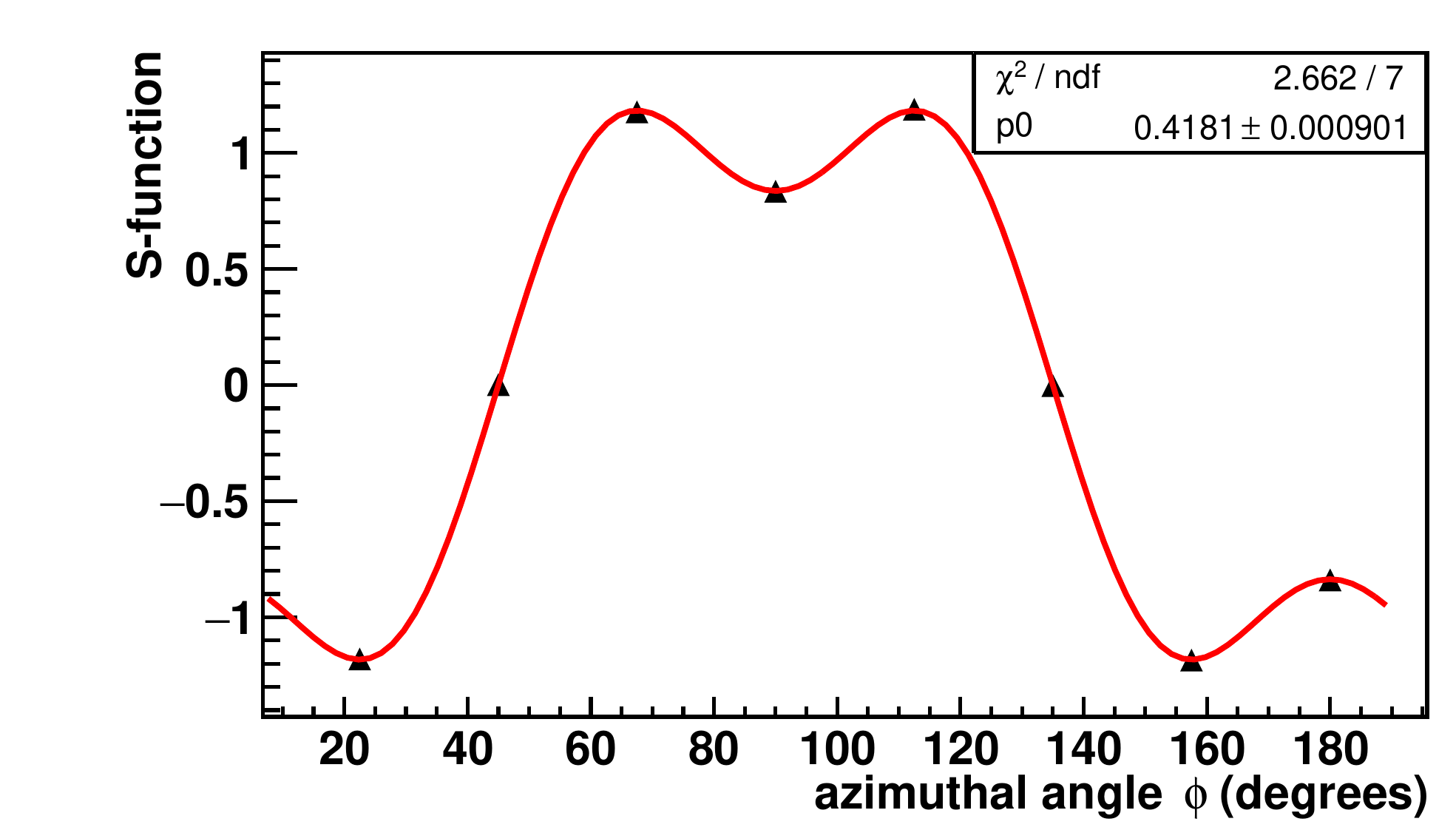}
\includegraphics[width=.48\textwidth]{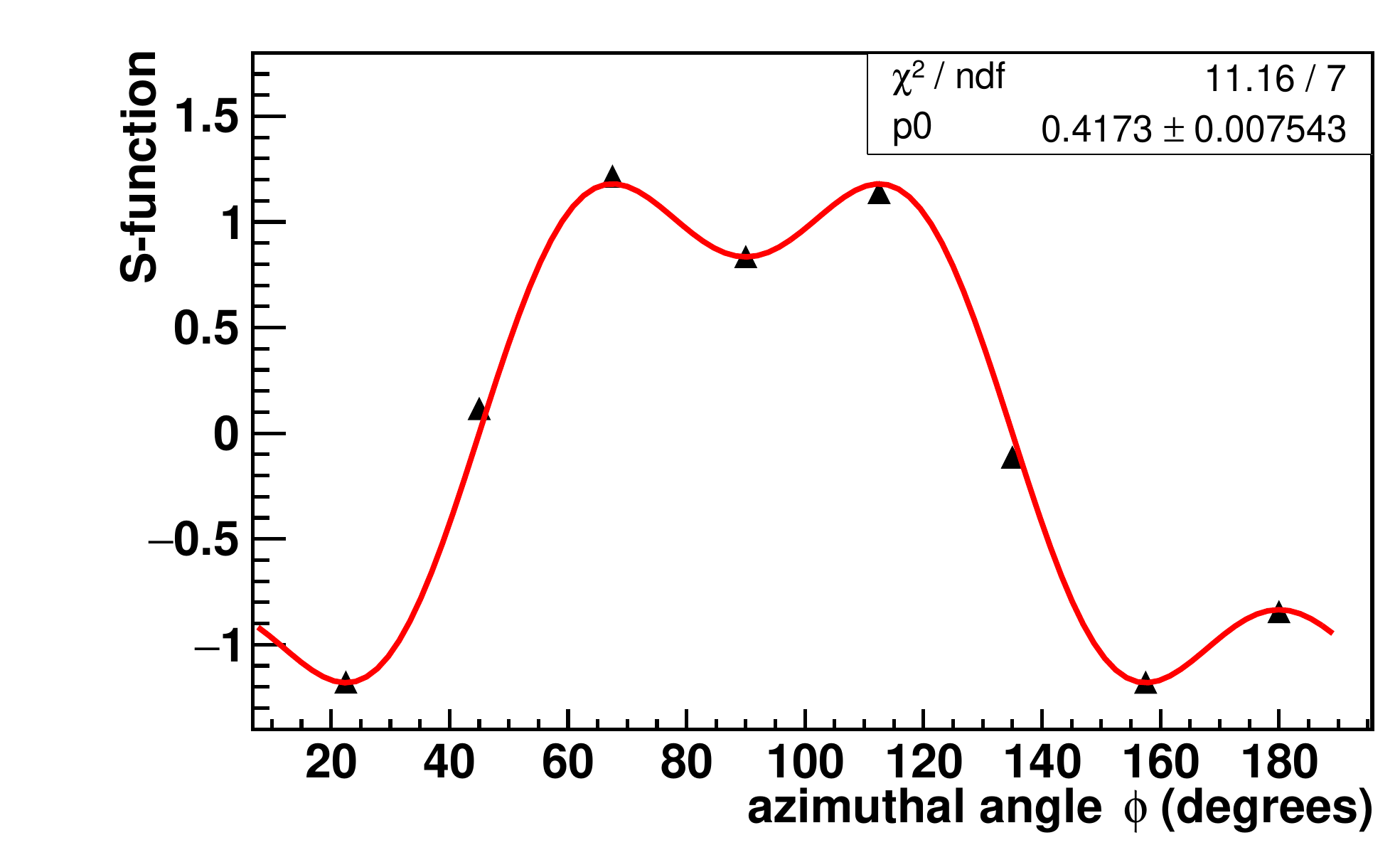}
\caption{\label{fig:i} Dependence of the S-function on the relative azimuthal angle between polarimeters for entangled (left) and decoherent photons of class "a" (right).  The solid line corresponds to the fitting function Eq.~\ref{eq:Sfunction}}
\label{fig:S-function}
\end{figure}

The measured $S-$functions for entangled and decoherent annihilation photons are presented in Fig.~\ref{fig:S-function}. In the latter case,  class ``a"  events with an energy deposition below 30 keV in the intermediate GAGG scatterer were selected to ensure the minimum difference from the kinematics of entangled photons. The fit of the experimental points by the theoretical function Eq.~\ref{eq:Sfunction} is shown by the solid line. The normalizing factor $p0$ is equal to the product of the analyzing powers of two polarimeters. Surprisingly, the behavior of both $S-$functions is almost identical regardless of the quantum state of the photons. Moreover, the normalizing factors are practically identical and coincide with the modulation factors estimated from the angular dependencies of the coincidence counts in the detectors of scattered photons, see Fig.~\ref{all_events}.

The identity of the $S$-functions indicates that the CHSH inequality, even corrected for the efficiency of Compton polarimeters, cannot be used to test the entanglement of annihilation photons.

\section*{Discussion}

In this work, we utilized two methods for studying entanglement in the Compton scattering of annihilation photons. The first method is based on the discovery by Bohm and Aharonov \cite{Bohm} of the relationship between strong angular correlations of scattered photons and entanglement.  According to them, a large ratio $R>2$ of the number of photon counts for the perpendicular and parallel orientations of the scattering planes indicates non-local correlations of annihilation photons.

The second approach involves the CHSH inequality, which is directly related to Bell's theorem. Osuch et al. \cite{Osuch} measured the $S$-function in this inequality and found that it agrees with quantum theory. Their correction of the $S$-function for the efficiency of Compton polarimeters confirmed the violation of the CHSH inequality.

The agreement between the theoretical and experimental results led to the conclusion that the system of two annihilation photons is entangled \cite{Caradonna}.  However, a recent paper by Hiesmayr and Moskal~\cite{Hiesmayr}  has shaken the belief. They found that the cross section for Compton scattering of annihilation photons, described by the Eq.~\ref{eq:prob}, is identical for entangled and mixed (separable)  state Eq.~\ref{eq:density_matrix}.  Since the cross section is of fundamental importance and determines all dependencies between scattered photons, the angular distributions and values of the $S$-function in the CHSH inequality must be the same for both quantum states.  Nevertheless, Caradonna et al.~\cite{Caradonna} disagreed with Hiesmayr and Moskal \cite{Hiesmayr}, confirming the results of Bohm and Aharonov \cite{Bohm}.

In our experiment, we compared Compton scattering of initial annihilation photons and decoherent pairs  prepared by preliminary  forward scattering of one of the initial photons before measurements in a polarimeter. The new results have revealed the same behavior of the angular distributions and $S$-functions for both quantum states. (The correlations of a backscattered photon with its pair are smaller, but we came to the conclusion that this state cannot be approximated by the Eq.~\ref{eq:density_matrix}.) Our measurements reveal that the entanglement of annihilation photons is still awaiting experimental confirmation using more advanced techniques.

Also, according to new experimental results, the similarity of angular correlations of scattered photons limits the use of quantum entanglement in the planned next generation  positron emission tomography (QE-PET). Most scattered background events cannot be rejected by the constraints on the angular distribution of scattered photon pairs.  At the same time, the random background obviously has no angular correlations and can be suppressed by applying appropriate kinematic cuts.

Perhaps the most intriguing discovery is the identity of the $S$-function in the CHSH inequality for the initial and prepared separable state of photon pairs, Eq.~\ref{eq:density_matrix}. However,  this is a consequence of the same Compton scattering cross section for both quantum states, as stated in ref.~\cite{Hiesmayr}. The equivalence of these cross sections raises the question of the universality of Bell's theorem for testing the entanglement and nonlocality of quantum theory. This issue  has been theoretically debated for a long time \cite{Khrennikov, Landau}. For example, a very recent discussion \cite{Griffiths2020, Lambare2021, Griffiths2021}  predicts a violation of CHSH inequality in a completely local system of neon $^{21}{\rm Ne}$ atoms with spin 3/2. The results presented provide an experimental basis for these theoretical discussions.

\bibliography{sample}

\section*{Acknowledgements}

The authors are grateful to B. Zhuikov for fabricating the $^{22}{\rm Na}$ radioactive source of positrons and to D. Levkov, M. Libanov, and A. Panin for useful discussions.

\end{document}